\documentclass[aps,twocolumn,showpacs,floatfix]{revtex4-1}
\usepackage{graphicx}
\usepackage{dcolumn}
\usepackage{bm}
\usepackage{epsf}
\usepackage{subfigure}
\usepackage{epstopdf}
\usepackage{amsmath}
\usepackage{amssymb}

\newcommand{\mean}[1]{\langle #1 \rangle}

\newcommand{\e}{{\rm e}}
\makeatletter
\def\rightharpoonupfill@{%
  \arrowfill@\relbar\relbar\rightharpoonup}
\def\leftharpoondownfill@{%
  \arrowfill@\leftharpoondown\relbar\relbar}

\newcommand{\xrightleftharpoons}[2][]{\mathrel{%
\raise.22ex\hbox{%
$\ext@arrow 3095\rightharpoonupfill@{\phantom{#1}}{#2}$}%
\setbox0=\hbox{%
$\ext@arrow 0359\leftharpoondownfill@{#1}{\phantom{#2}}$}%
\kern-\wd0 \lower.22ex\box0}%
}
\makeatother 
\begin{document}

\title{The Three Faces of the Second Law: I. Master Equation Formulation}
\author{Massimiliano Esposito}
\affiliation{Center for Nonlinear Phenomena and Complex Systems, Universit\'e Libre de Bruxelles, CP 231, Campus Plaine, B-1050 Brussels, Belgium.}
\author{Christian Van den Broeck}
\affiliation{Faculty of Sciences, Hasselt University, B-3590 Diepenbeek, Belgium.}

\date{\today}

\begin{abstract}
We propose a new formulation of stochastic thermodynamics for systems subjected to nonequilibrium constraints (i.e. broken detailed balance at steady state) and furthermore driven by external time-dependent forces. A splitting of the second law occurs in this description leading to three second law like relations. The general results are illustrated on specific solvable models. The present paper uses a master equation based approach.   
\end{abstract}

\pacs{05.70.Ln,05.40.-a}


\maketitle
\section{Introduction}\label{Intro}

The second law of thermodynamics specifies that the total entropy of an isolated macroscopic system cannot decrease in time. This statement applies to the stages of the evolution in which the entropy is well defined. For example, for a system in equilibrium at initial and final times, the final entropy will be larger than the initial one, even though the entropy may not be well defined during the intermediate evolution. However, it is often a very good approximation to assume that the system is in a state of local equilibrium, so that the entropy is well defined at any stage of the process. For example, linear irreversible thermodynamics is built on such an assumption, allowing the use of the Gibbs relation to define entropy locally in terms of the slow conserved quantities (for example  momentum, energy, and concentration of the constituents) \cite{GlansdorfPrigogine71,GrootMazur,Nicolis79}. The second law can then be reformulated as the non-negativity of the irreversible entropy production (EP) $\dot{S}_{i}(t) \geq 0$ \cite{PrigoThermo,Prigogine}:
\begin{eqnarray}
\dot{S}(t) = \dot{S}_e(t) + \dot{S}_{i}(t)  \;,
\label{Standard}
\end{eqnarray}
where $\dot{S}(t)$ is the entropy change of the considered subpart, and $\dot{S}_e(t)$ is the entropy flow to the environment. In some cases, the environment is idealized as being one or more reservoirs without internal dissipation, so that their entropy change is equal to minus the exchange term: $\dot{S}_r(t)=-\dot{S}_e(t)$. For a single heat reservoir, this entropy exchange is given by the energy inflow divided by its temperature. In the sequel, we will be mainly interested in this situation, with the subsystem of interest, henceforth called the system, in contact with an environment consisting of one or more ideal reservoirs. The irreversible EP in this system is then equal to the total EP $ \dot{S}_{tot}(t) \equiv \dot{S}_{i}(t) \geq 0$. 

In  more recent developments, \cite{Schnakenberg, Sekimoto98, LuoVdBNicolis84, Evans93, Evans94, Gallavotti95, Gallavotti95b, Jarzynski97, Jarzynski97b, Kurchan98, Crooks98, Lebowitz99, Hatano99, Maes99, Crooks99, Crooks00, HatanoSasa01, MaesNetocny03, Maes03, Seifert05, CleurenVdBKawaiPRL06, AndrieuxGaspard07a, AndrieuxGaspard07b, EspositoHarbola07PRE, Ge09, GaveauPRE09, EspositoVdBPRL10} it has been realized that one can formulate thermodynamics for small systems incorporating the effect of the fluctuations. These developments can be seen as the continuation of the pioneering work started by Onsager \cite{OnsagerReci1, OnsagerReci2, OnsagerMachlup1, OnsagerMachlup2}, with as intermediate steps the fluctuation dissipation theorem \cite{Callen51}, the theory of Gaussian stochastic processes and linear response \cite{FoxUhlenbeck70a, FoxUhlenbeck70b}, and Green-Kubo relations \cite{Kubo57, Kubo57b}. The essential ingredient is to guarantee the consistency of the statistical irreversible laws on the system dynamics with the reversibility of the equilibrium reservoirs statistics. In this new thermodynamics, also called stochastic thermodynamics \cite{VanDenBroeckST86, SeifertST08}, the system is described by a probability distribution $p_m$ evolving according to a Markovian master equation. The exchange of energy (heat) or particles with the environment and the other thermodynamic quantities associated to the system states $m$ become stochastic variables. The rates associated to each reservoirs satisfy the property of local detailed balance reminiscent of the fact that they always remain at equilibrium. The system entropy is defined using the Shannon expression $S=-\Sigma_m p_m \ln p_m$ and entropy balance equations of the usual form can be derived via the identification of a non-negative EP consistent with macroscopic nonequilibrium thermodynamics. A minimum EP theorem can also be proved \cite{Klein54, Schnakenberg, LuoVdBNicolis84, VanDenBroeckST86, MaesMEP07, MaesMEP07b}.

More recently, it was realized that one can study stochastic trajectory-dependent quantities. This is obviously the case for the energy, which is a well defined mechanical quantity even for single trajectories. For example, an explicit formulation of the first law, conservation of total energy, was given for the stochastic trajectory of a Langevin equation by Sekimoto \cite{Sekimoto98}, see also \cite{ParrondoEspagnol96}. The stochastic exchange of energy with idealized reservoirs also allows to identify the stochastic entropy flow into these reservoirs, and to study its statistical properties. This led to the discovery of the celebrated fluctuation theorem: the probability distribution for the cumulated change in stochastic reservoir entropy $\Delta s_r$ for a nonequilibrium steady state obeys the symmetry relation $P(\Delta s_r)/P(-\Delta s_r)\sim \exp(\Delta s_r)$ for asymptotically large times. We will use henceforth the notation lower case $s$ for entropic contributions associated to a given stochastic trajectory, in contrast to ensemble average entropies denoted by a capital letter $S$. This result was first proven for thermostated systems (the main trust of this work being however the development of equilibrium-like statistical mechanical concepts for dissipative systems \cite{Evans93, Evans94, Gallavotti95, Gallavotti95b}), followed by derivations for Langevin and master equations \cite{Kurchan98, Lebowitz99}, and for Hamiltonian dynamics \cite{MaesNetocny03, Maes03}. The asymptotic nature of the result was linked to the large deviation properties of the characteristic function, and in particular to those of the currents. Implications include Onsager symmetry and beyond \cite{AndrieuxGaspard04} and universal features of efficiency of thermal machines at maximum power \cite{VandenBroeckPRL05, EspositoPRL09}. Very much in the spirit of the fluctuation theorem, Jarzynski \cite{Jarzynski97,Jarzynski97b} and Crooks \cite{Crooks98,Crooks99,Crooks00} obtained the work theorem, see also \cite{VandenBroeck07,VandenBroeck08,VdBNJP09}. They find that the probability distribution for the work $w$ performed on a driven system, initially in canonical equilibrium at a specific inverse temperature $\beta^{-1}$, obeys the relation $P(w)/\bar{P}(-w)= \exp\{\beta(w-\Delta F)\}$. The over-bar corresponds to the probability distribution for the time-reversed experiment. If one assumes that, at the end of the driving, the system relaxes back to equilibrium (at inverse temperature $\beta^{-1}$) then $\beta(w-\Delta F)$, the so-called dissipated heat, is equal to the change in total entropy $\Delta s_{tot}$ of system plus reservoir, so that the work theorem becomes a fluctuation theorem for the change in total entropy $\Delta s_{tot}$, $P(\Delta s_{tot})/\bar{P}(-\Delta s_{tot})= \exp(\Delta s_{tot})$ \cite{MaesNetocny03,Seifert05}. Note that this result is valid for all times. The previous asymptotic fluctuation theorem can in fact be seen as a special case if one assumes that, aside from an initial transient, the steady state can be maintained for long enough times by appropriate driving, so that $\Delta s_{tot}=\Delta s+\Delta s_r \approx  \Delta s_r$. The focus on the asymptotic form and the accent on large deviation properties is in our opinion a somewhat misleading representation of the fluctuation theorem. This form arises from the neglect of the system's entropy (the so-called boundary term), for which no readily acceptable interpretation was deemed to exist at the time of the first formulations of the fluctuation theorem. Furthermore, the validity of the theorem is typically compromised for a system with unbounded energy \cite{Farago1,Farago2,Zon,Zonb,Maes06}, which is of course more the rule rather than the exception. We note however that current fluctuation theorems \cite{AndrieuxGaspard07a,AndrieuxGaspard07b} do require the long time limit since currents are related to the $\Delta s_r$ part of the entropy. 

A further advance consisted in the formulation of the equivalent of the second law at the stochastic level. This required the identification of a trajectory-dependent system entropy. Even though the idea is well known in information theory, where $-\ln p_m$ is the surprise at observing outcome $m$ when its probability is $p_m$, it took some time before Seifert \cite{Seifert05} identified this quantity  as the appropriate stochastic system entropy, $s(t)=-\ln p_{m(t)}(t)$. Note that it depends on the actual state $m(t)$ of the considered trajectory at the considered time, as well as on the probability for this state, which itself is in general time-dependent. By taking this term into account, the asymptotic fluctuation theorem could be replaced by a fluctuation theorem for the total entropy, which is valid for all times, just as the work theorem of Jarzynski and Crooks. Oono and Paniconi \cite{Oono98} discussed an alternative way of splitting the EP, by introducing the excess entropy and housekeeping heat. This led to the formulation of two other fluctuation theorems, namely one derived by Hatano and Sassa \cite{HatanoSasa01} (for system entropy plus excess entropy) and one by Speck and Seifert \cite{Seifert06} (for the housekeeping heat).

To close this introductory discussion, we turn to a recent development \cite{EspositoVdBPRL10}, see also \cite{EspositoHarbola07PRE, ChernyakJarzynski06, Harris07,Ge09}, which provides a clarifying and unifying approach of the various fluctuation and work theorems. The total stochastic EP $\Delta s_{tot}$ is the sum of two constitutive parts, namely a so-called adiabatic $\Delta s_{a}$ and nonadiabatic $\Delta s_{na}$ contribution. Each of these contributions correspond to the two basic ways that a system can be brought out of equilibrium: by applying steady nonequilibrium constraints (adiabatic contribution) or by driving (nonadiabatic contribution). Note that the term "adiabatic" is used here, not in its meaning referring to the absence of heat exchange, but in the meaning of instantaneous relaxation to the steady state. The crucial point is the observation that each of these contributions obeys a separate fluctuation relation, namely \cite{EspositoVdBPRL10}:
\begin{eqnarray}
&&\hspace{0.6cm}\frac{P(\Delta s_{tot})}{{\bar{P}}(-\Delta s_{tot})} = e^{\Delta s_{tot}}  \label{FT1} \\
&&\hspace{-1.3cm} \frac{P(\Delta s_{na})}{{\bar{P}}^+(-\Delta s_{na})} = e^{\Delta s_{na}} \ \ \;, \ \
\frac{P(\Delta s_{a})}{{P}^+(-\Delta s_{a})} = e^{\Delta s_{a}} \label{FT3}.
\end{eqnarray}
The superscript $+$ refers to the adjoint dynamics (also called dual or reversal \cite{Crooks00, ChernyakJarzynski06}). The aforementioned fluctuations theorems of Hatano Sassa and by Speck and Seifert are special case of the fluctuation theorem for $\Delta s_{na}$ and  $\Delta s_{a}$, respectively. To stress the special status of these theorems, we notice that they arise because of the two available operations to gauge the amount of time-symmetry breaking, namely time-reversal of the driving (overbar: $-$) and the time-reversal of the nonequilibrium boundary  conditions (superscript: $+$). Applying each of them separately, or both leads to the three different contributions for the EP.

The above fluctuation theorems imply that the total, adiabatic and nonadiabatic entropy changes have to be non-negative, each taken separately:
\begin{eqnarray}
&&\hspace{0.6cm} \Delta S_{tot}=\mean{\Delta s_{tot}} \ge 0\ \\
&&\hspace{-1.3cm} \Delta S_{na}=\mean{\Delta s_{na}} \ge 0 \ \ \;, \ \  
\Delta S_{a} =\mean{\Delta s_{a}} \ge 0.
\end{eqnarray}

This suggests that the second law can in fact be split in two. There are thus three faces to the second law: the increase of the average total entropy, the increase of the average adiabatic entropy and the increase of the average nonadiabatic entropy.
Our purpose here is to clarify and document further the physical properties and the meaning of this remarkable result.
In this paper we will focus on the implications for a description in terms of a master equation. The next paper \cite{EspoVdB10_Db} deals with the corresponding results for Langevin and Fokker Planck dynamics.

\section{Master Equation}

\subsection{Entropy balance}

We first review and extend the entropy balance equation derived previously for a Markovian process \cite{Schnakenberg, LuoVdBNicolis84, Gaspard04b, EspositoHarbola07PRE}. 
Our starting point is the following master equation:
\begin{eqnarray}
\dot{p}_m(t) = \sum_{m'} W_{m,m'} \;p_{m'}(t) , \label{ME}
\end{eqnarray}
where the rate matrix satisfies 
\begin{eqnarray}
\sum_{m} W_{m,m'}=0 . \label{ProbCons}
\end{eqnarray}
The transitions between states $m$ can be due to different mechanism $\nu$. Furthermore, these rates can be time-dependent via a control variable $\lambda$. We thus have: 
\begin{eqnarray}
W_{m,m'} = W_{m,m'}(\lambda_t) = \sum_{\nu} W_{m,m'}^{(\nu)}(\lambda_t) . \label{DiffMech}
\end{eqnarray}
For rates that are ``frozen" at the values $W_{m,m'}^{(\nu)}(\lambda)$, there is a corresponding stationary distribution, $p_{m}^{st}(\lambda)$, which we suppose to be unique (i.e. the rate matrix is irreducible) and which will always eventually be reached by the system. It is given by the normalized right eigenvector of zero eigenvalue of the transition matrix:
\begin{eqnarray}
\sum_{m'} W_{m,m'}(\lambda) p_{m'}^{st}(\lambda) = 0 . \label{SSDist}
\end{eqnarray}

Let us now investigate the time dependence of the system's Shannon entropy (Boltzmann's constant $k_B=1$):
\begin{eqnarray}
S(t) =- \sum_{m}  p_{m}(t)  \ln p_{m}(t). \label{ShannonEntropy}
\end{eqnarray}
Using (\ref{ME}) and (\ref{ProbCons}) and omitting for compactness of notation the dependence of $p_m$ on $t$ and of $W$ on $\lambda_t$ we find:
\begin{eqnarray}
\dot{S}(t) &=&- \sum_{m} \dot{p}_{m}\ln p_{m} =- \sum_{m,m',\nu} W^{(\nu)}_{m,m'} \;p_{m'}\ln \frac{p_{m}}{p_{m'}}\\
&=&\frac{1}{2}\sum_{m,m',\nu} \{W^{(\nu)}_{m,m'} \;p_{m'}-W^{(\nu)}_{m',m} \;p_{m}\}\ln \frac{p_{m'}}{p_{m}}\nonumber\\
&=&\frac{1}{2}\sum_{m,m',\nu} \{W^{(\nu)}_{m,m'} \;p_{m'}-W^{(\nu)}_{m',m} \;p_{m}\}\ln \frac{W_{m,m'}^{(\nu)}\;p_{m'}}{W_{m',m}^{(\nu)}\;p_{m}} \nonumber \\
&&+\frac{1}{2}\sum_{m,m',\nu} \{W^{(\nu)}_{m,m'} \;p_{m'}-W^{(\nu)}_{m',m}\; p_{m}\}\ln \frac{W_{m',m}^{(\nu)}}{W_{m,m'}^{(\nu)}} \nonumber
\end{eqnarray}
It is revealing to introduce the fluxes $J_{m,m'}^{(\nu)}$ and corresponding forces $X_{m,m'}^{(\nu)}$:
\begin{eqnarray}
&&\hspace{-0.5cm} J_{m,m'}^{(\nu)}(t) = W^{(\nu)}_{m,m'}(\lambda_t) \;p_{m'}(t)-W^{(\nu)}_{m',m}(\lambda_t) \;p_{m}(t), \label{Flux} \\
&&\hspace{-0.5cm} X_{m,m'}^{(\nu)}(t) = \ln \frac{W_{m,m'}^{(\nu)}(\lambda_t)\;p_{m'}(t)}{W_{m',m}^{(\nu)}(\lambda_t)\;p_{m}(t)}. \label{Force}
\end{eqnarray}
We can rewrite the master equation (\ref{ME}) as
\begin{eqnarray}
\dot{p}_m(t) = \sum_{m',\nu} J_{m,m'}^{(\nu)}(t) = \sum_{m'} J_{m,m'}(t) , \label{ME2} 
\end{eqnarray}
where we defined 
\begin{eqnarray}
J_{m,m'}(t) = \sum_{\nu} J_{m,m'}^{(\nu)}(t). \label{TotFluxDef}
\end{eqnarray}
The system's EP can thus be rewritten under the familiar form of irreversible thermodynamics
\begin{eqnarray}
\dot{S}(t) = \dot{S}_{e}(t)+ \dot{S}_{i}(t). \label{SecondLaw}
\end{eqnarray}
The quantity:
\begin{eqnarray}
\dot{S}_{e}(t) &=& \frac{1}{2} \sum_{m,m',\nu} J_{m,m'}^{(\nu)}(t) \ln \frac{W_{m',m}^{(\nu)}(\lambda_t)}{W_{m,m'}^{(\nu)}(\lambda_t)} \label{EntropyFlow}\\
&=& \sum_{m,m',\nu} W_{m,m'}^{(\nu)}(\lambda_t) p_{m'}(t) \ln \frac{W_{m',m}^{(\nu)}(\lambda_t)}{W_{m,m'}^{(\nu)}(\lambda_t)} \nonumber
\end{eqnarray}
is the entropy flow and the positive quantity
\begin{eqnarray}
\dot{S}_{i}(t) &=& \sum_{m,m',\nu} W_{m,m'}^{(\nu)}(\lambda_t) p_{m'}(t) \ln  \frac{W_{m,m'}^{(\nu)}(\lambda_t) p_{m'}(t)}{W_{m',m}^{(\nu)}(\lambda_t)p_m(t)} \nonumber \\
&=& \frac{1}{2} \sum_{m,m',\nu} J_{m,m'}^{(\nu)}(t)\;X_{m,m'}^{(\nu)}(t) \geq 0,  \label{StandBilinear}
\end{eqnarray}
is identified as the EP. The latter is zero if and only if the condition of detailed balance is satisfied:   
\begin{eqnarray}
W_{m,m'}^{(\nu)}(\lambda) p_{m'} = W_{m',m}^{(\nu)}(\lambda) p_{m}. \label{DBalance}
\end{eqnarray}

We note an important property of the EP. If all the relevant processes $\nu$ causing transitions between states $m$ are not correctly identified (for example if one only identifies a sub-class of these processes) the EP will be underestimated. Indeed, using the log-sum inequality (cf. theorem 2.7.1 of \cite{CoverThomas}), it follows that (see also \cite{VandenBroeck07,VandenBroeck08,VdBNJP09}):
\begin{eqnarray}
\dot{S}_{i}(t) &=& \sum_{m,m',\nu} W_{m,m'}^{(\nu)}(\lambda_t) p_{m'}(t) \ln  \frac{W_{m,m'}^{(\nu)}(\lambda_t) p_{m'}(t)}{W_{m',m}^{(\nu)}(\lambda_t)p_m(t)} \nonumber \\
&\geq& \sum_{m,m'} W_{m,m'}(\lambda_t) p_{m'}(t) \ln  \frac{W_{m,m'}(\lambda_t) p_{m'}(t)}{W_{m',m}(\lambda_t)p_m(t)} \geq 0 \nonumber \\ \label{EPCoarseGrained}
\end{eqnarray}

\subsection{Thermodynamic interpretation}

The above derivations and statements can be viewed as purely mathematical in nature and can be applied to any system described by a master equation. The connection with physics is made by associating to each mechanism $\nu$ responsible for the transitions between system states a group of variables each separately at it own equilibrium or in other word idealized reservoirs with well defined thermodynamic variables (e.g. temperature or chemical potential). The transitions between states $m$ due to different mechanism $\nu$ can for example correspond to exchange of heat with different reservoirs, or the change in number of particles due to different chemical reactions. As a result the transition rates associated to a given mechanism $\nu$ need to obey the condition of local detailed balance
\begin{eqnarray}
W_{m,m'}^{(\nu)}(\lambda_t) p_{m'}^{eq}(\lambda_t,\nu) = W_{m',m}^{(\nu)}(\lambda_t) p_{m}^{eq}(\lambda_t,\nu) , \label{eqDist}
\end{eqnarray}
where the equilibrium distribution $p^{eq}(\lambda_t,\nu)$ is the stationary distribution that would be reached by the system if only a single mechanism $\nu$ were present and for the frozen value of the control variable $\lambda=\lambda_t$. In case multiple mechanism are present, each reservoir tries unsuccessfully to impose its equilibrium distribution on the system resulting in a stationary distribution that does not satisfy the detailed balance condition (\ref{DBalance}) except if their thermodynamical properties are identical, making the distinction between the various mechanism useless. As we have seen in (\ref{EPCoarseGrained}), a incorrect identification of the various reservoirs would underestimate the EP and could lead to believe that a system is at equilibrium while it it not \cite{LuoVdBNicolis84}. The assumption that the same expressions for the transition probabilities in (\ref{eqDist}) can be used when the control parameter becomes time-dependent is based on an assumption that the idealized reservoirs relax infinitely fast to their equilibrium compared to the timescales of the system dynamics. 

By an argument of physical consistency, it follows that $\dot{S}_{i}(t)$ is also equal to the total EP $\dot{S}_{tot}(t)$ in system plus environment. Indeed, since one implicitly assumes that the environment remains at equilibrium at all times using (\ref{eqDist}), it does not have an internal EP of its own. Otherwise, the above description in terms of the system alone is not complete, and one needs to include the description of the irreversible processes taking place in the environment. For the same reason, the entropy flow is equal to minus the entropy increase in the reservoir $\dot{S}_{e}(t)=-\dot{S}_{r}(t)$. The microscopic origin of these relations has been recently clarified \cite{EspoLindVdBNJP10}.

In order to make the thermodynamic interpretation of the stochastic dynamics transparent, we now explicitly evaluate the various entropies for multiple heat reservoirs satisfying local detailed balance with respect to the canonical equilibrium distribution at the inverse temperature of the corresponding reservoir, i.e. (\ref{eqDist}) becomes:
\begin{eqnarray}
\frac{W_{m,m'}^{(\nu)}(\lambda)}{W_{m',m}^{(\nu)}(\lambda)}=\exp{\{-\beta^{(\nu)} \big( \epsilon_m(\lambda)-\epsilon_{m'}(\lambda)\big)\}} , \label{rateKMS}
\end{eqnarray}
where $\epsilon_m(\lambda)$ is the energy of the system when in the state $m$ for the value $\lambda$ of the control variable. We note that in order to make the connection between the system Shannon entropy (\ref{ShannonEntropy}) and the true thermodynamic entropy, one might have to add a contribution $\sum_m S_m p_m(t)$ to the entropy, where $S_m$ is the entropy of each level $m$ \cite{AndrieuxPNAS08,EspositoJSM10,GaveauPRE09}. However, for simplicity, we now assume that the index $m$ refers to the non-degenerate microscopic state of the system (i.e., $S_m=0$). It immediately follows from (\ref{rateKMS}) that the entropy flow (\ref{EntropyFlow}) takes on the familiar thermodynamic form of heat flux over temperature:
\begin{eqnarray}
\dot{S}_{e}(t)= \sum_{\nu} \beta^{(\nu)} \dot{Q}^{(\nu)}(t), \label{EntropyFlowExpl}
\end{eqnarray}
where the heat from the $\nu$ reservoir is given by
\begin{eqnarray}
\dot{Q}^{(\nu)}(t) = \sum_{m,m'} J_{m,m'}^{(\nu)}(t) \epsilon_{m'}(\lambda) \label{Heat} .
\end{eqnarray}
Since the system energy is obviously given by:
\begin{eqnarray}
E(t) = \sum_{m} \epsilon_{m}(\lambda) p_m(t) , \label{AverEnergy}
\end{eqnarray}
we find from the first principle of thermodynamics (energy conservation),
\begin{eqnarray}
\dot{E}(t) = \dot{W}(t) + \sum_{\nu} \dot{Q}^{(\nu)}(t) , \label{FirstPrinc}
\end{eqnarray}
that the work is given by:
\begin{eqnarray}
\dot{W}(t) = \sum_{m} \dot{\epsilon}_{m}(\lambda) p_m . \label{Work}
\end{eqnarray}
This illustrates that the local detailed balance conditions with the reservoirs (\ref{rateKMS}) leads to a proper formulation of he first (\ref{FirstPrinc}) and second (\ref{SecondLaw}) principle of thermodynamics. We note that it is possible to include particles exchanges with the reservoirs \cite{EspositoPRL09}. In the latter case, work can be nonzero even in absence of driving. \\

We should finally mention that the system could also be driven out of equilibrium by a nonconservative force instead of different reservoirs. In this case, the local detailed balance condition (\ref{eqDist}) is not satisfied and even in presence of a single reservoir, the detailed balance condition (\ref{DBalance}) will not be satisfied at steady state.  

\subsection{Adiabatic and nonadiabatic entropy balance}
 
We next note that the force $X$ (\ref{Force}) can be split in an adiabatic contribution $A$ and a nonadiabatic contribution $N$:
\begin{eqnarray}
X_{m,m'}^{(\nu)}(t) &=& A_{m,m'}^{(\nu)}(\lambda_t) + N_{m,m'}(t) \label{ForceSplitt}\\
A_{m,m'}^{(\nu)}(\lambda_t) &=& \ln \frac{W_{m,m'}^{(\nu)}(\lambda_t)\;p^{st}_{m'}(\lambda_t)}{W_{m',m}^{(\nu)}(\lambda_t)\;p^{st}_{m}(\lambda_t)} \label{AdiabForce}\\
N_{m,m'}(t) &=& \ln \frac{p^{st}_{m}(\lambda_t)\;p_{m'}(t)}{p^{st}_{m'}(\lambda_t)\;p_{m}(t)}. \label{NAdiabForce}
\end{eqnarray}

The total EP can thus also be split into an adiabatic and nonadiabatic contribution: 
\begin{eqnarray}
\dot{S}_{i}(t) \equiv \dot{S}_{tot}(t) = \dot{S}_{a}(t)+ \dot{S}_{na}(t) \label{EPsplitt}
\end{eqnarray}
with:
\begin{eqnarray}
\dot{S}_{a}(t) &=& \frac{1}{2} \sum_{m,m',\nu}J_{m,m'}^{(\nu)}(t)\;A_{m,m'}^{(\nu)}(\lambda_t) \geq 0 . \label{EPa} \\
&=& \sum_{m,m',\nu} W_{m,m'}^{(\nu)}(\lambda_t) p_{m'}(t) \ln  \frac{W_{m,m'}^{(\nu)}(\lambda_t) p_{m'}^{st}(\lambda_t)}{W_{m',m}^{(\nu)}(\lambda_t)p_m^{st}(\lambda_t)} \nonumber \\
\dot{S}_{na}(t) &=& \frac{1}{2} \sum_{m,m'}J_{m,m'}(t)\;N_{m,m'}(t)\geq 0. \label{EPana} \\
&=& - \sum_{m} \dot{p}_{m}(t)\ln \frac{p_{m}(t)}{p^{st}_{m}(\lambda_t)} \nonumber 
\end{eqnarray}
We make a number of remarks. First and most importantly, both $\dot{S}_{a}(t)$ and $\dot{S}_{na}(t)$ are non-negative EPs. This follows from Jensen's inequality, $-\ln x \geq 1-x$ for $x>0$, together with (\ref{ProbCons}) and (\ref{SSDist}). The non-negativity of these quantities is in agreement with the fact that the trajectory entropies ${s}_{a}$ and ${s}_{na}$ obey detailed fluctuation theorems \cite{EspositoVdBPRL10}. Second, it is clear that the nonadiabatic thermodynamic force, and hence the nonadiabatic EP, is zero in the adiabatic limit $p_{m}(t) \rightarrow p^{st}_{m}(\lambda_t)$. This will for example be the case when the relaxation to the steady state is extremely fast, and in particular faster than the time-scale of the driving $\lambda_t$. This observation justifies a posteriori the name given to each contribution. Third, we note that one does not need to identify the separate mechanisms $\nu$ by which the transition between states takes place for the evaluation of nonadiabatic EP, or its corresponding fluxes and forces. It is function only of the coarse grained transition probabilities $W=\sum_{\nu} W^{\nu}$. Finally, we note for a system subjected to a nonconservative force and in contact with a single reservoir, the adiabatic EP is the Housekeeping heat (divided by the temperature of the reservoir) \cite{Oono98,HatanoSasa01,SpeckSeifert05,Harris07}.

The fact that there are two contributions to the total entropy production, which are separately non-negative, suggests that the second law can be ``split in two". An elegant way to do so  is by the introduction of the so-called excess entropy change \cite{Oono98,HatanoSasa01}:
\begin{eqnarray}
\dot{S}_{ex}(t) &=& \frac{1}{2} \sum_{m,m'} J_{m,m'}(t) \ln \frac{p_{m}^{st}(\lambda_t)}{p_{m'}^{st}(\lambda_t)} \label{EPExcess} \\
&=& \sum_{m} \dot{p}_{m}(t) \ln p^{st}_{m}(\lambda_t) \nonumber .
\end{eqnarray}
One easily verifies that:
\begin{eqnarray}
\dot{S}(t) &=& -\dot{S}_{ex}(t) + \dot{S}_{na}(t) \label{newSecondLawI} \\
\dot{S}_r(t) &=& \dot{S}_{ex}(t) + \dot{S}_{a}(t) \label{newSecondLawII}.
\end{eqnarray}
Written under this form, we see that  the changes in system and reservoir entropy  both have a structure similar to the original second law (\ref{SecondLaw}): they consist of the sum of a reversible and an irreversible term. By summing these two relations we recover (\ref{EPsplitt}). The relations (\ref{EPsplitt}), (\ref{newSecondLawI}) and (\ref{newSecondLawII}) thus represent  the three faces of the second law.

\section{Specific class of transformations}

Additional comments can be made when considering the specific classes of transformation discussed below. To proceed, it is useful to split the nonadiabatic entropy (\ref{EPana}) into a ``boundary" and a ``driving" part \cite{EspositoHarbola07PRE}:
\begin{eqnarray}
\dot{S}_{na}(t) = \dot{S}_{b}(t) + \dot{S}_{d}(t) \label{EPna2},
\end{eqnarray}
where 
\begin{eqnarray}
\dot{S}_{b}(t) &=& - \frac{d}{dt} \bigg( \sum_{m} p_m(t) \ln \frac{p_m(t)}{p^{st}_{m}(\lambda_t)} \bigg) \label{Boundary} \\
\dot{S}_{d}(t) &=& - \sum_{m} \frac{p_m(t)}{p^{st}_{m}(\lambda_t)} \dot{p}^{st}_{m}(\lambda_t) \label{Driving}.
\end{eqnarray}
The boundary contribution only depends on the initial and final distribution of the considered transformation while the driving part is only non-zero when the external perturbation evolves in time. 

\subsection{Transient relaxation to steady state}

The system is supposed to be in an arbitrary state when the external driving is switched off, say  at $t=0$. For $t>0$, we have that $\lambda_t=\lambda$ is time-independent,  implying $\dot{S}_{d}(t)=0$, and therefore:
\begin{eqnarray}
\dot{S}_{na}(t) = \dot{S}_{b}(t) = - \dot{H}(t).\label{EPna2}
\end{eqnarray}
Here, $H(t)$ is the relative entropy (or Kullback-Leibler entropy) \cite{CoverThomas} between the actual and the steady state distribution:
\begin{eqnarray}
H(t) = D(p(t)||p^{st}(\lambda)) = \sum_{m} p_{m}(t) \ln \frac{p_{m}(t)}{p^{st}_{m}(\lambda)} \geq 0.
\end{eqnarray}
We conclude that $H(t)$ is a Lyapunov function, decreasing monotonically in time until the probability distribution reaches its steady state value. This proof of convergence to the steady state is well know \cite{KampenB97}, but its connection to the nonadiabatic EP has never been pointed out. The above result can also be viewed as a generalization of the minimum EP principle (MEP) \cite{LuoVdBNicolis84}: the nonadiabatic entropy assumes its minimum value equal to zero at the steady state, but the latter need not be close to equilibrium. 

\subsection{Transitions between steady states}

We consider a system that is initially in a steady state $p_m(0)=p^{st}_m(\lambda_{t_i})$. Then somewhere between $t_i$ and $t_f$ the time dependent force changes and after an asymptotically long time $T$ the system has reached its new steady state distribution $p_m(T)=p^{st}_m(\lambda_{t_f})$. Let us consider the total nonadiabatic entropy change during the time interval $T$. The boundary term is zero and the driving term is nonzero between $t_i$ and $t_f$. The relation (\ref{newSecondLawI}) becomes the second law of steady state thermodynamics \cite{Oono98,HatanoSasa01}:
\begin{eqnarray}
\Delta S_{na}(T)=\Delta S_{d}(T)=\Delta S_{ex}(T) + \Delta S^{st}(T) \geq 0 \label{LawISStransitions},
\end{eqnarray}
where $\Delta S^{st}(T)$ is the system entropy change between the initial and final steady state.  

\subsection{Time dependent cycles}

We consider a system that is  subjected to a time-periodic perturbation with period  $T$. After an initial transient, the probability distribution and the various EP's  will also become time-periodic functions with the same period. Because of continuity,  we have $p_m(0)=p_m(T)$. Furthermore $p^{st}_m(\lambda_0)=p^{st}_m(\lambda_T)$. Hence both the boundary term along a cycle as well as the system entropy change over a period are zero, $\Delta S(T)=\Delta S_{b}(T)=0$, and the relation (\ref{newSecondLawI}) becomes
\begin{eqnarray}
\Delta S_{na}(T) = \Delta S_{ex}(T) = \Delta S_{d}(T) \geq 0 \label{LawICycles}.
\end{eqnarray}
We also have, using (\ref{SecondLaw}), that:
\begin{eqnarray}
\Delta S_{tot}(T) = \Delta S_{r}(T) = \Delta S_{ex}(T) + \Delta S_{a}(T) \geq 0
\end{eqnarray}


\subsection{Perturbing the steady state}

Finally, we investigate how the adiabatic and nonadiabatic EP change upon applying a perturbation around the steady state distribution:
\begin{eqnarray}
p_{m}(t) = p_{m}^{st}(\lambda) + \delta p_m(t) \label{DistribPert}.
\end{eqnarray}
Clearly, the flux (\ref{Flux}) can be split as: 
\begin{eqnarray}
J_{m,m'}^{(\nu)}(t) = J_{m,m'}^{(\nu)}(\lambda) + \delta J_{m,m'}^{(\nu)}(t),
\end{eqnarray}
where
\begin{eqnarray}
J_{m,m'}^{(\nu)}(\lambda) &=& W^{(\nu)}_{m,m'}(\lambda) \;p_{m'}^{st}(\lambda)-W^{(\nu)}_{m',m}(\lambda) \;p_{m}^{st}(\lambda) \nonumber \\ \label{aFlux} \\
\delta J_{m,m'}^{(\nu)}(t) &=& W_{m,m'}^{(\nu)}(\lambda) \; \delta p_{m'}(t)
-W_{m',m}^{(\nu)}(\lambda) \; \delta p_{m}(t) \nonumber \\ \label{naFlux}
\end{eqnarray}

The adiabatic contribution to the EP can be written as the sum of two contributions
\begin{eqnarray}
\dot{S}_{a}(t) = \dot{S}_{a}(\lambda) + \delta \dot{S}_{a}(t) \geq 0 \label{EPaSplit} ,
\end{eqnarray}
with:
\begin{eqnarray}
\dot{S}_{a}(\lambda) &=& \frac{1}{2} \sum_{m,m',\nu} J_{m,m'}^{(\nu)}(\lambda) \; A_{m,m'}^{(\nu)}(\lambda) \geq 0 \label{EPa1}\\
\delta \dot{S}_{a}(t) &=& \frac{1}{2} \sum_{m,m',\nu} \delta J_{m,m'}^{(\nu)}(t) \; A_{m,m'}^{(\nu)}(\lambda). \label{EPa2} 
\end{eqnarray}

We now turn to the nonadiabatic EP which using (\ref{EPana}) and (\ref{DistribPert}) now reads:
\begin{eqnarray}
\dot{S}_{na}(t) = - \sum_{m} \delta \dot{p}_{m}(t) \ln \frac{p_{m}(t)}{p_{m}^{st}(\lambda)} \geq 0 .\label{DeltaEPna} 
\end{eqnarray}
In analogy to the total EP [which can be written in a bilinear form in term of flux and forces (\ref{StandBilinear})], we can write the nonadiabatic EP as a bilinear form of the nonadiabatic flux and force as:
\begin{eqnarray}
\dot{S}_{na}(t) = - \sum_{m,m'} \delta J_{m,m'} N_{m,m'} \geq 0 ,\label{DeltaEPnabis} 
\end{eqnarray}
noting however that:
\begin{eqnarray}
N_{m,m'}(t)=\frac{ \delta p_{m'}}{p^{st}_{m'}}-\frac{ \delta p_{m}}{p^{st}_{m}} +O(\delta^2 p). \label{NAdiabForceexp}
\end{eqnarray}

We conclude that the adiabatic EP has a constant (zero order) and first order terms  in the deviation $\delta p$ around the stationary state $ p^{st}_m$, while the nonadiabatic EP is of second order. In general, since $\delta \dot{S}_{a}(t)$ can be negative, we can find situations where $\dot{S}_{a}(t)$ and even $\dot{S}_{i}(t)$ are smaller then $\dot{S}_{a}(\lambda)$. This is the well known result that the MEP for the total EP does not hold around nonequilibrium steady states but does around equilibrium steady states since $A_{m,m'}^{(\nu)}(\lambda)=0$ \cite{Schnakenberg, LuoVdBNicolis84}.

\section{Applications}

\subsection{Two level system}

We consider a system with two levels, $m=1,2$, with $p_m$ the probability to find the system in level $m$. Due to conservation of total probability, the master equation (\ref{ME}) can be reduced to a single differential equation for the probability $p=p_2=1-p_1$ to be in the level $2$: 
\begin{eqnarray}
\dot{p}(t) = - \gamma(\lambda_t) p(t) + W_{21}(\lambda_t), \label{ME2level}
\end{eqnarray}
with $\gamma(\lambda) = W_{21}(\lambda) + W_{12}(\lambda)$. The steady state solution is $p^{st}(\lambda)=W_{21}(\lambda)/\gamma(\lambda)$. 

We first present the general solution for a periodic perturbation, with a period $T$, without  specifying the  form of the rates at this stage. We focus on the long-time regime, where all transients have disappeared, and the time behavior of  ${p}(t)$ itself is periodic with period $T$.  We need to solve  (\ref{ME2level}) subject to  periodic boundary conditions $p(0)=p(T)$.  For simplicity, we focus on the case of  a piece-wise constant  perturbation. Hence, the  driving period  consists of two regimes $I$ and $II$, namely $ \lambda_t=\lambda_I$  for $0 \leq t < t_I$, and $\lambda_t=\lambda_{II}$ for $t_I \leq t < T $. The solution of (\ref{ME2level}) reads ($p^{st}_{I}=p^{st}(\lambda_I)$, $p^{st}_{II}=p^{st}(\lambda_{II})$):
\begin{eqnarray}
&&\hspace{-0.7cm} 0 \leq t < t_I: \  \  p(t) = \big( p(T)-p^{st}_I \big) \e^{-\gamma_I t} + p^{st}_I \label{ME2GenSolA} \\
&&\hspace{-0.7cm} t_I \leq t < T: \  \  p(t) = \big( p(t_I)-p^{st}_{II} \big) \e^{-\gamma_{II} (t-t_I)} + p^{st}_{II} \label{ME2GenSolB} .
\end{eqnarray}
Using the matching condition (continuity of $p$) at the transitions between  regions $I$  and $II$, we find that:
\begin{eqnarray}
&& p(t_I) =  \frac{p^{st}_{I} (\e^{\gamma_{I} t_I}-1) - p^{st}_{II} (\e^{-\gamma_{II} (T-t_I)}-1)}{\e^{\gamma_{I}t_I}-\e^{-\gamma_{II}(T-t_I)}} \label{PeriodCond1}\\
&& p(T) = \frac{p^{st}_I (\e^{-\gamma_I t_I}-1) - p^{st}_{II} (\e^{\gamma_{II} (T-t_I)}-1)}{\e^{-\gamma_{I}t_I}-\e^{\gamma_{II}(T-t_I)}} . \label{PeriodCond2}
\end{eqnarray}
(\ref{PeriodCond1}) and (\ref{PeriodCond2}) with (\ref{ME2GenSolA}) and (\ref{ME2GenSolB}) provide the exact and explicit long-time solution of (\ref{ME2level}) under piece-wise constant periodic driving.

\begin{figure}[h]
\centering
\begin{tabular}{c@{\hspace{0.5cm}}c}
\hspace{0.35cm}
\rotatebox{0}{\scalebox{0.35}{\includegraphics{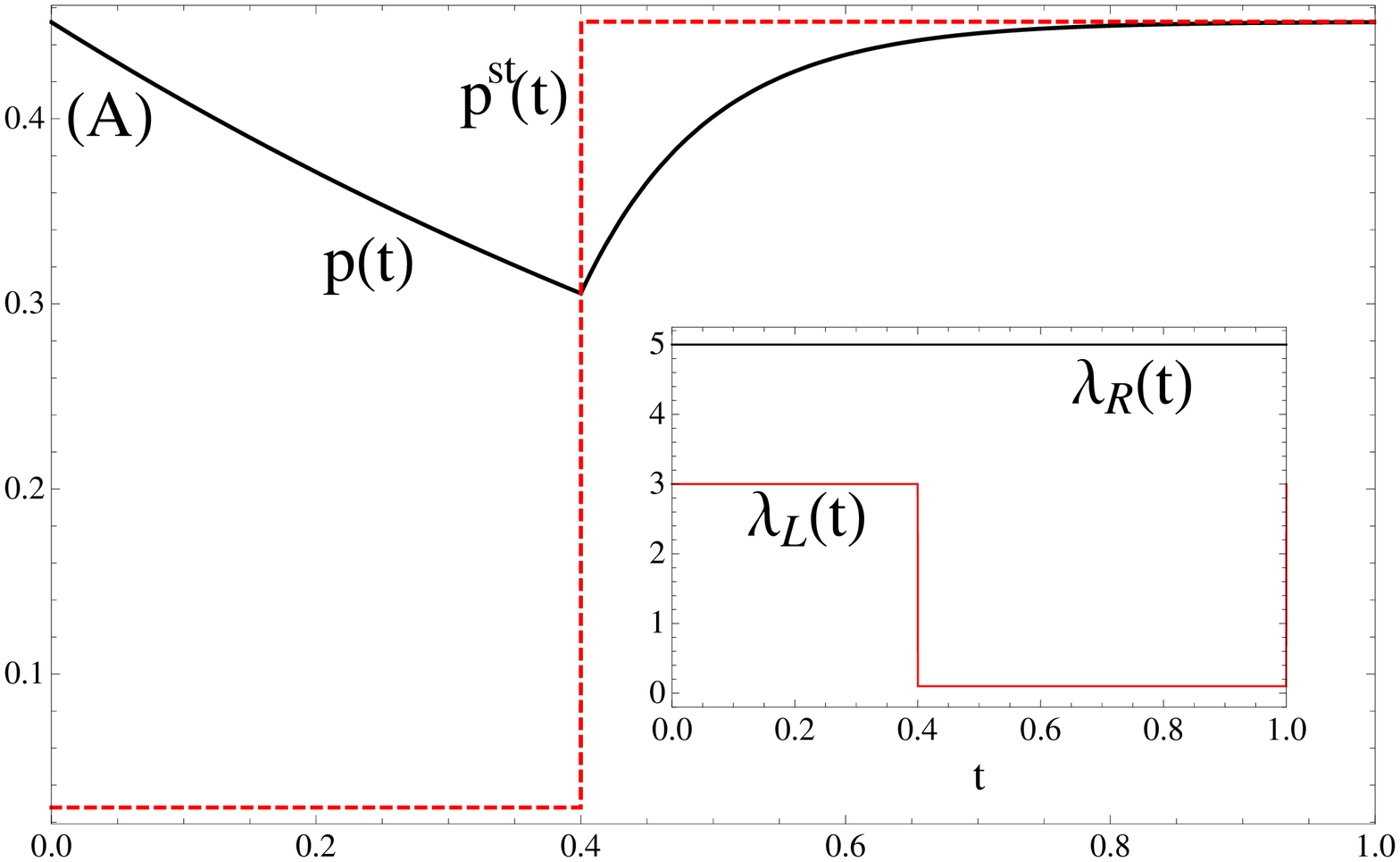}}} \vspace{-0.35cm}\\ \hspace{0.28cm}
\rotatebox{0}{\scalebox{0.35}{\includegraphics{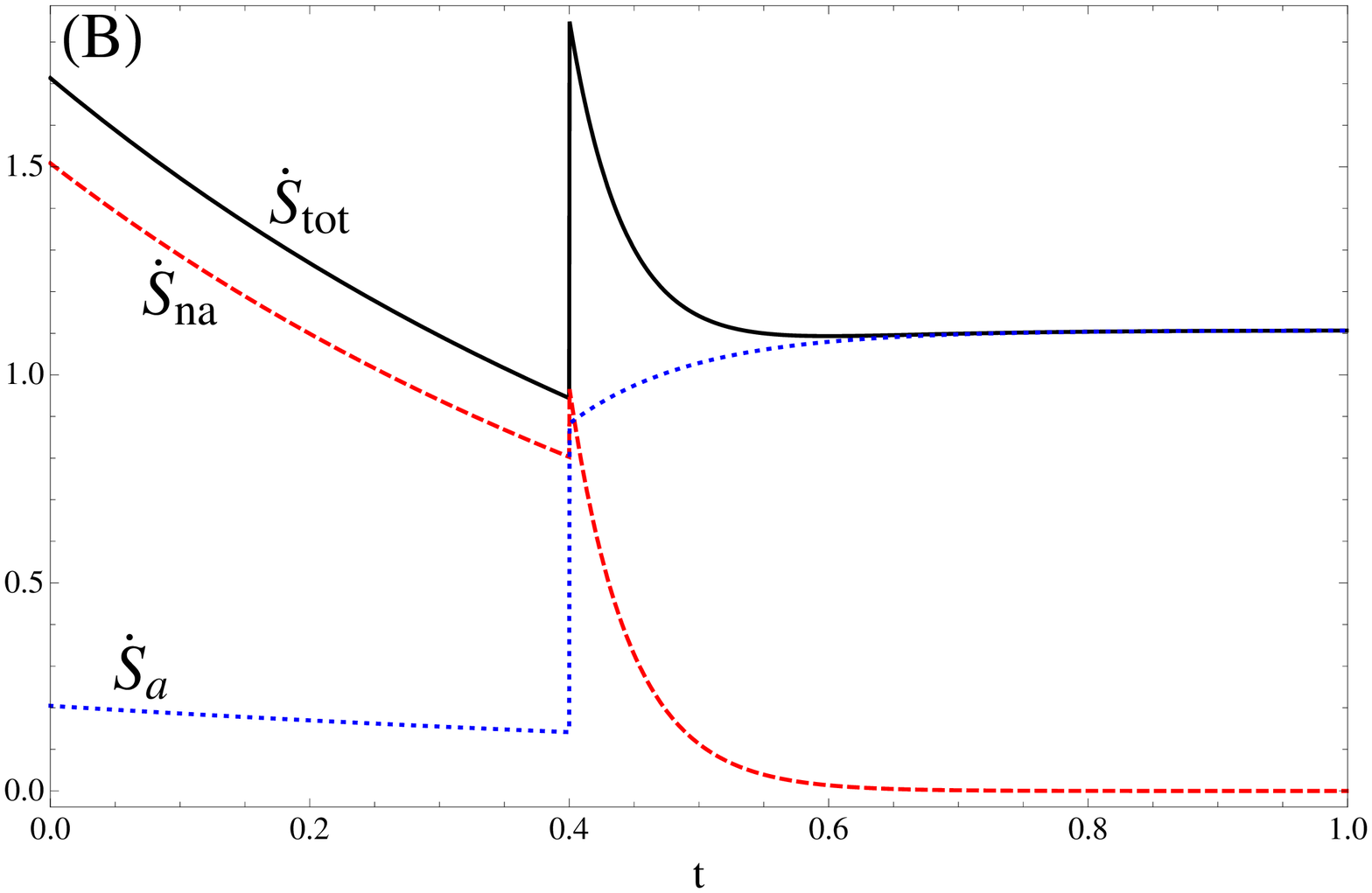}}}
\end{tabular}
\caption{(Color online) (A) Actual probability distribution along the cycle and stationary solution corresponding to the instantaneous values of the driving displayed in the inset. (B) The total, adiabatic and nonadiabatic entropy production along the cycle. The Bose rates (\ref{Bose}) are used with $\Gamma=0.5$. Also $T=1$ and $t_I=0.4$.}
\label{plot1}
\end{figure}

We now turn to the entropies. We consider the case of  two different reservoirs $\nu=L,R$, with corresponding transition rates $W=\sum_{\nu} W^{(\nu)}$. The entropies (\ref{EPsplitt}), (\ref{EPa}) and (\ref{EPana}) become
\begin{eqnarray}
&&\hspace{-0.7cm} \dot{S}_{tot} = \sum_{\nu} \big( W_{12}^{(\nu)} p - W_{21}^{(\nu)} ( 1-p ) \big) \ln \frac{W_{12}^{(\nu)} p}{W_{21}^{(\nu)} (1-p)} \label{EPtot2lev} \\
&&\hspace{-0.7cm} \dot{S}_{a} = \sum_{\nu} \big( W_{12}^{(\nu)} p - W_{21}^{(\nu)} (1-p) \big) \ln \frac{W_{12}^{(\nu)} p^{st}}{W_{21}^{(\nu)} (1-p^{st})} \label{EPa2lev} \\
&&\hspace{-0.7cm} \dot{S}_{na} = \dot{p}(t) \ln \frac{(1-p) p^{st}}{(1-p^{st}) p} \label{EPna2lev} 
\end{eqnarray}

As an illustration, we consider  two physical models that are described by the above two-level master equation. The first is a single electron quantum level in contact with a left ($L$) or right ($R$) electronic reservoirs ($\nu=L,R$). The level can thus be empty, $1-p$, or filled, $p$, and transitions between these two states correspond to the exchange of an electron with one of the two reservoirs. The rates are given by the Fermi golden rule rates for each of the reservoirs \cite{HarbolaEsposito06}
\begin{eqnarray}
W_{21}^{(\nu)} = \Gamma f(\lambda_{\nu}) \  \ , \  \ W_{21}^{(\nu)} = \Gamma \big(1-f(\lambda_{\nu})\big) \label{Fermi},
\end{eqnarray}
where $f(x)=(\exp{\{x\}}+1)^{-1}$. The thermodynamics of this model has been discussed in Refs. \cite{EspoLindVdB_EPL09_Dot,EspoKawLindVdB_PRE_10}. The time dependence of the rates can result from either the external control of the energy of the level (e.g. with an electric field) or from the control of the reservoirs chemical potentials. 

Our second model represents two level atom interacting with a left ($L$) and right ($R$) reservoir of thermal light. The rates describing the transitions between the levels are then given by:
\begin{eqnarray}
W_{21}^{(\nu)}=\Gamma n(\lambda_{\nu}) \  \ , \  \ W_{21}^{(\nu)}=\Gamma \big(1+n(\lambda_{\nu})\big) \label{Bose},
\end{eqnarray}
where $n(x)=(\exp{\{x\}}-1)^{-1}$. The time dependence of the rates can either come from an external control of the energy spacing between the levels or, which is less realistic, from the control of the temperature of the reservoirs. 

\begin{figure}[h]
\centering
\begin{tabular}{c@{\hspace{0.5cm}}c}
\hspace{0.35cm}
\rotatebox{0}{\scalebox{0.35}{\includegraphics{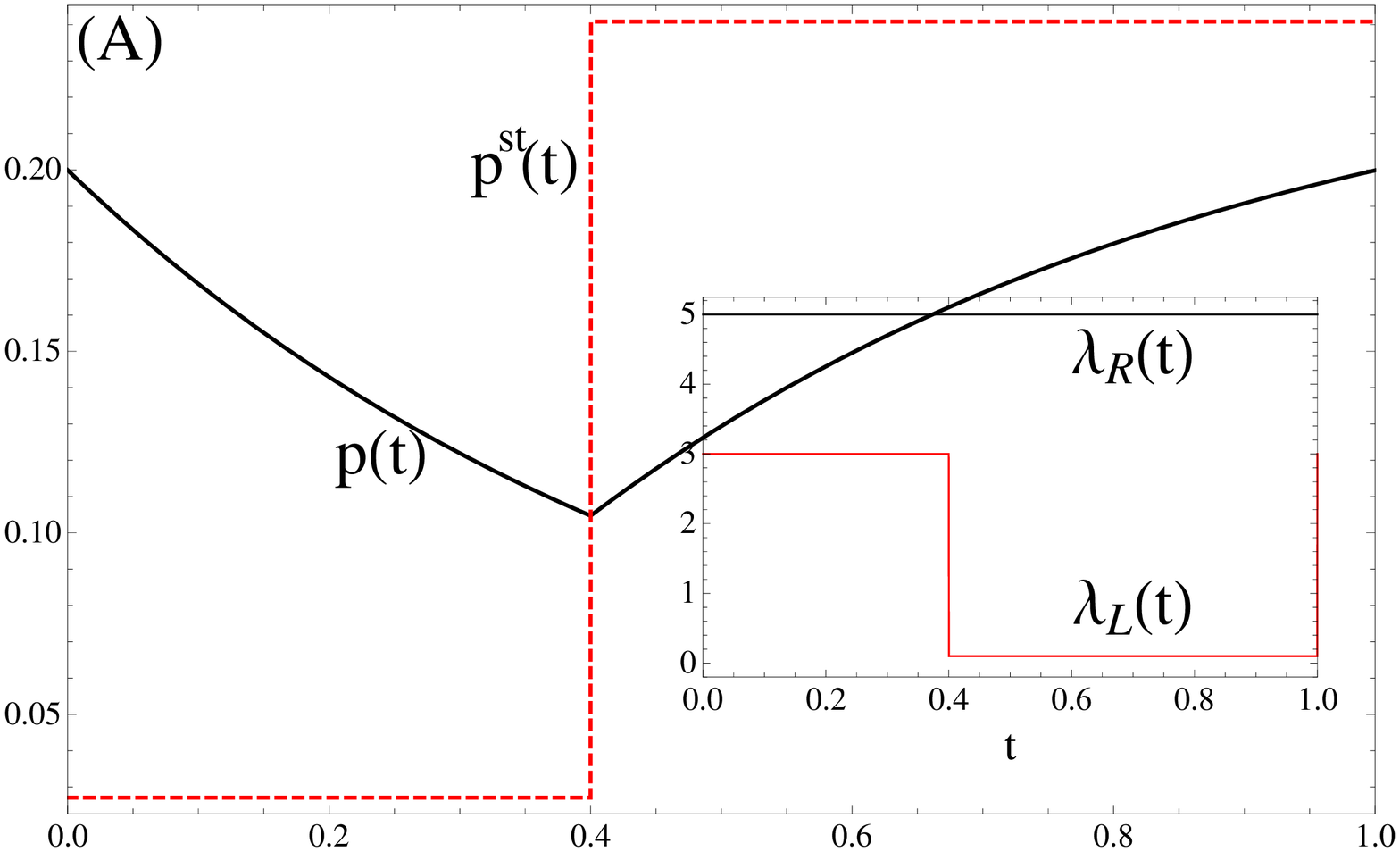}}} \vspace{-0.35cm}\\ \hspace{0.35cm}
\rotatebox{0}{\scalebox{0.35}{\includegraphics{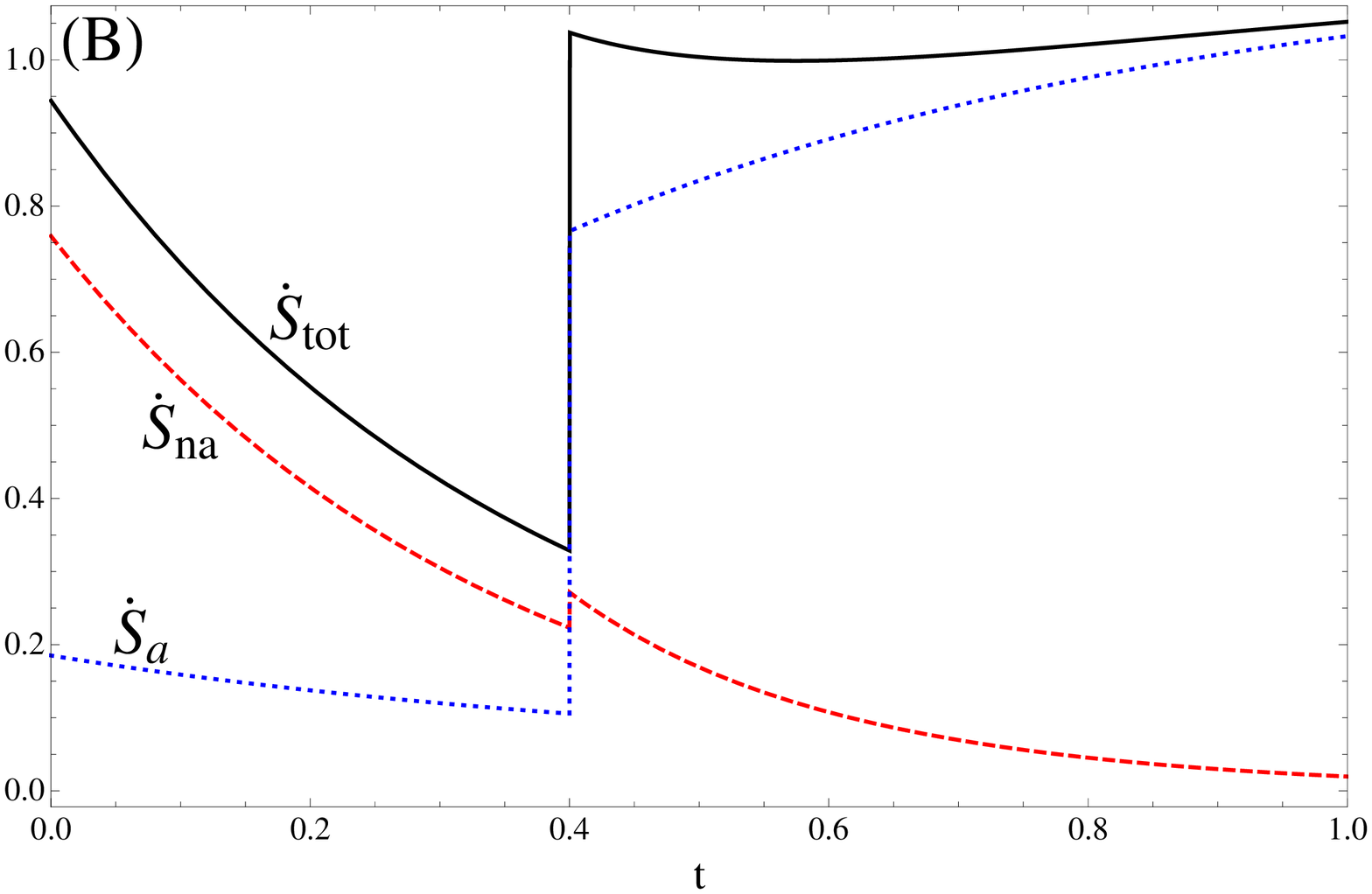}}}
\end{tabular}
\caption{(Color online) Same as Fig. \ref{plot1}, but using the Fermi rates (\ref{Fermi}) instead of the Bose rates.}
\label{plot2}
\end{figure}

The probability distribution as well as the various entropies around the cycle can be analytically evaluated. For the purpose of illustration, we reproduce the typical behavior  in  Figs. (\ref{plot1}) and (\ref{plot2}). 

\subsection{Chemical reaction model}

Consider the following chemical reaction:
\begin{eqnarray}
A^{(L)} \xrightleftharpoons[k_M^{(L)}]{k_A^{(L)}}  M \xrightleftharpoons[k_A^{(R)}]{k_M^{(R)}} A^{(R)}
\end{eqnarray}
Denoting by $m$ the number of particles of species $M$ and using $\nu=L,R$, we find that the corresponding transition rates are given by 
\begin{eqnarray}
W^{(\nu)}_{m-1,m} = k_M^{(\nu)} m \ \ , \ \ W^{(\nu)}_{m+1,m} = k_A^{(\nu)} A^{(\nu)}(t).
\end{eqnarray}
The concentrations of $A^{(L)}(t)$ and $A^{(R)}(t)$ are externally controlled in a time dependent manner.
Rescaling time by $k_M^{(L)}+k_M^{(R)}$, we can redefine the rates as
\begin{eqnarray}
W^{(\nu)}_{m-1,m} = \alpha^{(\nu)} m \ \ , \ \ W^{(\nu)}_{m+1,m} = \lambda^{(\nu)}_t. \label{RateChem1}
\end{eqnarray}
where we introducted
\begin{eqnarray}
\alpha^{(\nu)} = \frac{k_M^{(\nu)}}{k_M^{(L)}+k_M^{(R)}} \ \ , \ \  \lambda^{(\nu)}_t = \frac{k_A^{(\nu)} A^{(\nu)}(t)}{k_M^{(L)}+k_M^{(R)}}. \label{RateChem2}
\end{eqnarray}
The resulting master equation reads ($m=0,1,...$ with convention $p_{-1}=0$) \cite{KampenB97}:
\begin{eqnarray}
\dot{p}_m &=& \sum_{\nu} \{ \alpha^{(\nu)} (m+1) p_{m+1} + \lambda^{(\nu)}_t p_{m-1} \nonumber \\ && - (\lambda^{(\nu)}_t + \alpha^{(\nu)} m) p_{m} \}  \label{MEqChem} \\  
&=& (m+1) p_{m+1} + \lambda_t p_{m-1} - (\lambda_t + m) p_{m} \nonumber.
\end{eqnarray}
From first to second line, we used $\sum_{\nu} \lambda^{(\nu)}_t = \lambda_t$ and $\sum_{\nu} \alpha^{(\nu)} =1$.
Eq. (\ref{MEqChem}) can be most easily solved by switching to the generating function:
\begin{eqnarray}
F(s,t) = \sum\limits_{m=0}^{\infty} s^m  p_m(t). \label{GenFctChem}
\end{eqnarray}
Using (\ref{MEqChem}) and (\ref{GenFctChem}), one finds:
\begin{eqnarray}
\partial_t F(s,t) = (s-1) \lambda_t F(s,t) -(s-1) \partial_s F(s,t). \label{F}
\end{eqnarray}
For purpose of  illustration, we focus on a specific time-dependent solution of this equation, for which the explicit analytic expression can be obtained. Indeed, one readily verifies by inspection that:
\begin{eqnarray}
F(s,t)=e^{{{\bar{m}}}(t) (s-1)},
\end{eqnarray}
is an exact solution of (\ref{F}), propagating in time. As suggested by the notation, ${{\bar{m}}}(t)$ is the time dependent average:
\begin{eqnarray}
{\bar{m}}={\bar{m}}(t)=\sum\limits_{m=0}^{\infty} m p_m(t),\label{average}
\end{eqnarray}
which obeys the following equation:
\begin{eqnarray}
\dot{\bar{m}}(t)= \lambda_t-{\bar{m}}(t).\label{eqaverage}
\end{eqnarray}
Its exact time-dependent solution is given by:
\begin{eqnarray}
{\bar{m}}(t)= e^{-(t-t_0)} \{ {\bar{m}}(t_0) + \int\limits_{t_0}^{t} d\tau e^{(\tau-t_0)} \lambda_{\tau} \} .\label{averagesol}
\end{eqnarray}
We conclude that the propagating Poissonian distribution:
\begin{eqnarray}
p_m(t)=\frac{{\bar{m}}^m}{m!} e^{-{\bar{m}}}, \label{DistribChem}
\end{eqnarray}
with the average given by (\ref{averagesol}), is an exact time-dependent solution of the master equation. The corresponding frozen steady states are 
\begin{eqnarray}
p_m^{st}(t)=\frac{[{\bar{m}}^{st}]^m}{m!} e^{-{\bar{m}^{st}}}, \label{DistribChemStat}
\end{eqnarray}
with 
\begin{eqnarray}
{\bar{m}}^{st} = \lambda_t .\label{averagesolst}
\end{eqnarray}
We note, using (\ref{RateChem1}) and (\ref{averagesol}), that
\begin{eqnarray}
\frac{W^{(\nu)}_{m+1,m} p_m(t)}{W^{(\nu)}_{m,m+1} p_{m+1}(t)} = \frac{\lambda_t^{(\nu)}}{\alpha^{(\nu)} {\bar{m}}(t)} .\label{DBCondChem}
\end{eqnarray}
We see that at steady state, the detailed balance condition is not in general satisfied. Equilibrium is only attained if $\lambda_t^{(\nu)}=\alpha^{(\nu)} \lambda_t$. We thus have two mechanism bringing the system out of equilibrium, the breaking of detailed balance by the steady state and the time dependent driving from $\lambda_t^{(\nu)}$.

\begin{figure}[h]
\centering
\begin{tabular}{c@{\hspace{0.5cm}}c}
\hspace{0.35cm}
\rotatebox{0}{\scalebox{0.35}{\includegraphics{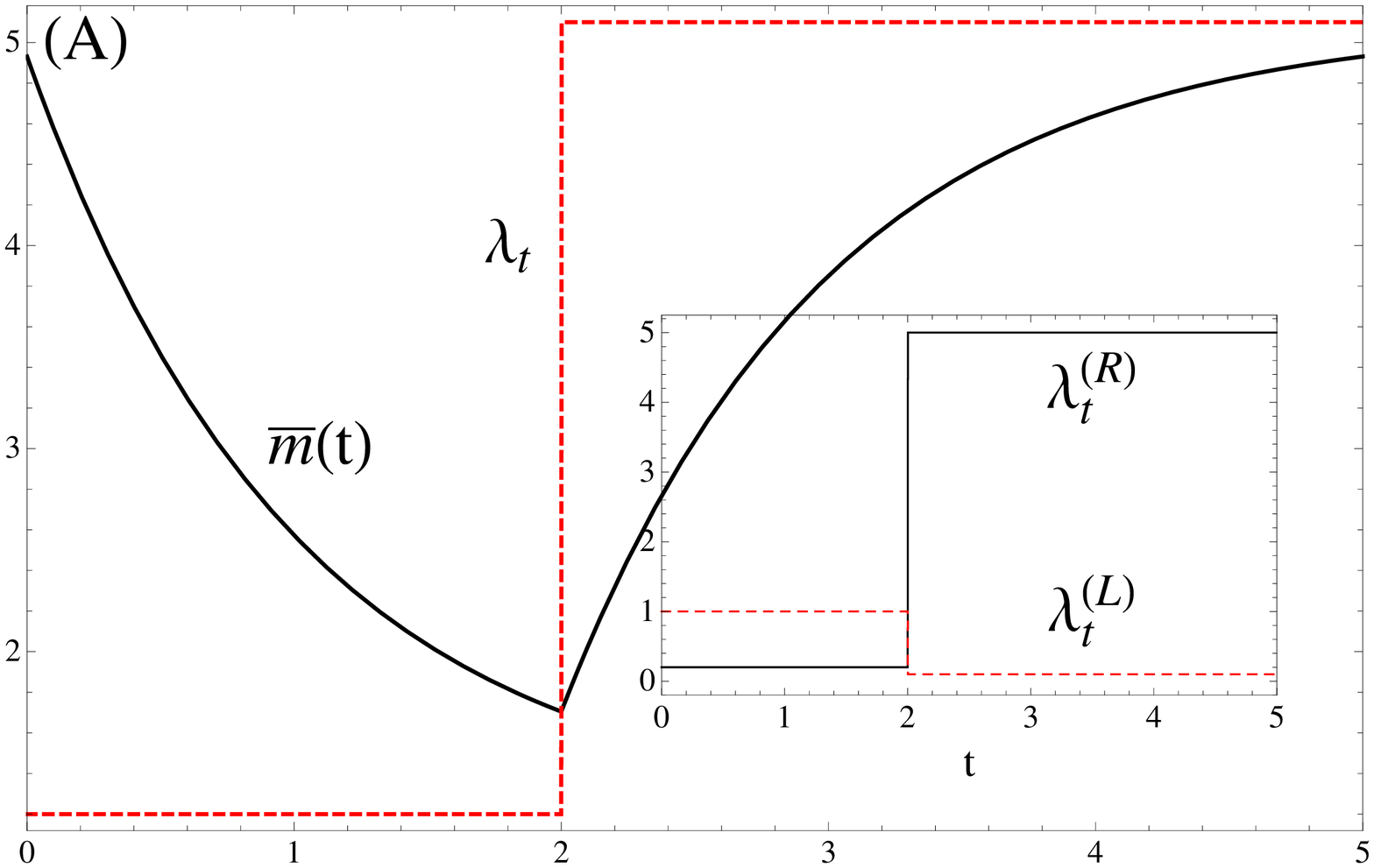}}} \vspace{-0.35cm}\\ \hspace{0.26cm}
\rotatebox{0}{\scalebox{0.35}{\includegraphics{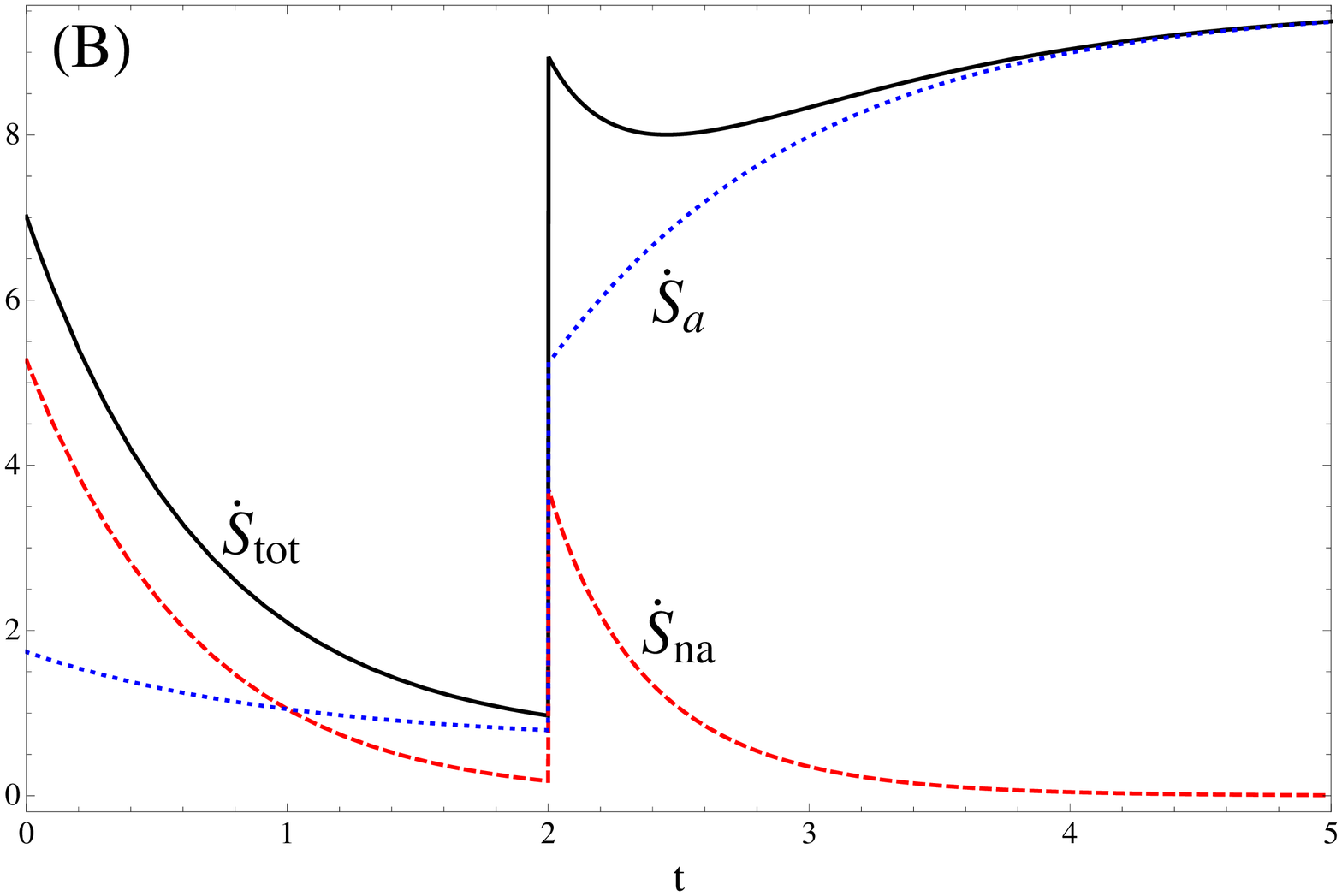}}}
\end{tabular}
\caption{(Color online) (A) Actual average $\bar{m}(t)$ and stationary average $\lambda_t$ along the cycle [they uniquely determine the actual and stationary probability distribution (\ref{DistribChem}) and (\ref{DistribChemStat})] corresponding to the instantaneous values of the driving displayed in the inset. (B) The total, adiabatic and nonadiabatic entropy production along the cycle. Also $T=5$ and $t_I=2$.}
\label{plot3}
\end{figure}

Using these analytical results, we can now obtain explicit results for the entropies (\ref{EPsplitt}), (\ref{EPa}) and (\ref{EPana}):
\begin{eqnarray}
{\dot{S}}_{tot}(t) &=& \sum_{\nu} (\alpha^{(\nu)} \bar{m}-\lambda_t^{(\nu)}) \ln \frac{\alpha^{(\nu)} \bar{m}}{\lambda_t^{(\nu)}} \label{stot} \\
{\dot{S}}_{a}(t) &=& \sum_{\nu} (\alpha^{(\nu)} \bar{m}-\lambda_t^{(\nu)}) \ln \frac{\alpha^{(\nu)} \bar{m}^{st}}{\lambda_t^{(\nu)}} \label{sac} \\
{{\dot{S}}_{na}}(t) &=& - \dot{\bar{m}} \ln \frac{\bar{m}}{\lambda_t}=({\bar{m}}-\lambda_t) \ln \frac{\bar{m}}{\lambda_t} ,\label{snac}
\end{eqnarray}
which are all positive. Not surprisingly, the nonadiabatic contribution vanishes in the adiabatic limit $\bar{m}=\lambda_t$. On the other hand, the adiabatic contribution vanishes under the instantaneous detailed balance condition $\lambda_t^{(\nu)}=\alpha^{(\nu)} \lambda_t$ (remembering that  $\lambda_t=\bar{m}^{st}$). 

To compare the results with the two-level model, we again assume that the system is subjected to a periodic piece-wise constant driving of period $T$. One finds:
\begin{eqnarray}
&&\hspace{-0.7cm} 0 \leq t < t_I: \  \  \bar{m}(t) = \e^{-t} \big( \bar{m}(T) + \lambda_I (\e^{t}-1) \big) \label{MEChemSolA} \\
&&\hspace{-0.7cm} t_I \leq t < T: \  \  \bar{m}(t) = \e^{-(t-t_0)} \big( \bar{m}(t_I) + \lambda_{II} (\e^{t-t_I}-1) \big) \nonumber,
\end{eqnarray}
where
\begin{eqnarray}
&& \bar{m}(t_I) = \frac{\lambda_I (\e^{-t_I}-1) + \lambda_{II} (\e^{-T}-\e^{-t_I})}{\e^{-T}-1} \label{ChemPeriodCond1}\\
&& \bar{m}(T) = \frac{\lambda_I (\e^{t_I}-1) + \lambda_{II} (\e^{T}-\e^{t_I})}{\e^{T}-1}. \label{ChemPeriodCond2}
\end{eqnarray}
The resulting behavior of the probability distribution and the various EP around the cycle are shown in Fig. (\ref{plot3}).

\section{Conclusions}

In this paper, we started by showing how to identify entropy and entropy production for a stochastic dynamics described by a Markovian master equations with time dependent rates. The key new element is that the entropy production can be ``split" in two parts, each satisfying a second law like relation. By assuming that the rates satisfy a local detailed balance condition, we have also shown that the dynamics provides a nonequilibrium thermodynamics description of the system. The thermodynamic implications of this new splitting remain to be properly understood. To progress in this direction we calculated the various entropies on a different exactly solvable models. In the companion paper \cite{EspoVdB10_Db}, we proceed similarly for a stochastic dynamics described by a Fokker-Plank equation. 

\section*{Acknowledgments}

M. E. is supported by the Belgian Federal Government (IAP project ``NOSY").

%

\end{document}